# MOSARIX: PROGRESS REPORT

*Abstract:* MOSARIX is a collaborative project between three research group in Sorbonne University to build a x-ray spectrometer (2-5 keV) portable to large scale facilities with high efficiency and good resolution. X-ray spectroscopy and coincidences experiment are planned. A prototype with a single HADP crystal with von Hamos geometry has been tested (resolution and efficiency). A fast time and position detector has been realized (patent and publication).We plan to build the spectrometer with 8 HADP (or 4) crystals under Helium atmosphere using a TimePix3 or a CCD camera.

MOSARIX is a project of an x-ray spectrometer in the tender x-ray domain (2 – 5 keV) with high efficiency, allowing performing x-ray emission and coincidences (or covariance mapping) experiments using synchrotron radiation, XFEL, the future installation SPIRAL2/GANIL or CRYRING/FAIR. It involves 2 groups at LCPMR (Francis PENENT and Marc SIMON) and one group at INSP (Dominique VERNHET). The coincidences/covariance measurements will be between x-ray photons and ions or electrons. It would be the first time for such coincidences with energy-resolved photons.

The spectrometer will be portable and will be brought to the different large-scale facilities.

MOSARIX is a multi-crystal HAPG von Hamos spectrometer optimized for the 2-5 keV photon energy range. Its resolving power $E/\Delta E$ will be 4000. It will be equipped with a fast time and position sensitive detection system, allowing performing coincidences, or with a CCD camera.

## I. Scientific case and some possible experiments

The accelerated development of x-ray sources, as 3[rd] generation synchrotrons (and recent upgrades) or free-electron lasers, has opened new opportunities to investigate new phenomena by means of photoelectron and Auger spectroscopy, electron-ion coincidence techniques and x-ray emission. However, several processes of high scientific interests are still hard to measure; some of them require the measurement of photons with high efficiency, high resolution and even sometimes in coincidence mode. This is the purpose of MOSARIX development.

As an example, we propose to revisit Resonance-Enhanced X-ray Multiple Ionization (REXMI)[1] with a significant amelioration of the detection of photons, i.e. measuring the photons not only with high efficiency and high resolution but also in coincidence with ions or electrons. This will allow accessing the involved intermediate states and obtaining a clearer image of the dynamic of the multiple ionization process. MOSARIX can also be used for the investigation of very low cross-section phenomena such as attosecond electron dynamics[2] and High-Energy Resolution Off-Resonant Spectroscopy (HEROS)[3,4].

X-ray spectroscopy has also proved to be a very powerful tool to investigate quantum dynamics in heavy ions collisions with matter of whatever nature, dilute or condensed[5–7]. A future installation SPIRAL2/GANIL in Caen will allow extracting excitation cross sections in fast ion - slow ion collisions. The coincidence measurement of photons (resolved in energy) with ions are of great importance to this project.

In the following, we particularly address three scientific projects that we prioritise for MOSARIX. However, the use of MOSARIX will not be limited to these projects. More challenging processes that have lower cross sections and require a coincidence detection of photons such as Radiative Auger processes or Resonance-Enhanced X-ray Multiple Ionization (REXMI) could be considered later on.

## High-energy resolution off-resonant spectroscopy

X-ray absorption spectroscopy is the technique of choice to probe the unoccupied states in a molecular system. On the other hand, high-energy resolution off-resonant spectroscopy (HEROS) allows measurement of scattered X-ray spectrum [3,4]. In this case, for an incident photon energy fixed below an absorption threshold, the X-ray emission spectrum is equivalent to a core-hole-broadening-free absorption spectrum. Only few off-resonance X-ray emission studies were successfully realized, particularly in gas-phase, because inelastic X-ray scattering is characterized by a very low cross section in this off-resonance regime. Most of the multi-crystals von Hamos spectrometers in this domain have been developed for the 5-10 keV energy range.

The work of Marchenko et al [2] on Inelastic X-ray Scattering on the molecule $CS_2$ molecule showed that for large so-called photon energy "detuning" with respect to the first absorption resonance, the spectra are sensitive to the electronic dynamics on an attosecond timescale. With very high detunings (100-500 eV), this technique will allow measuring electron dynamics on such a short timescale, complementary to pump-probe experiments using of attosecond HHG sources.

Combining Resonant Auger spectroscopy with RIXS measurements provide interesting information like the topology of the involved Potential Energy Surfaces and the lifetime of the Double Core Hole electronic states produced [8]. The main issues of HEROS are the very low cross section of the inelastic X- ray scattering (~10-3 compared to absorption cross section) and the need of high-energy-resolution detection systems. To overcome them, multi-crystal spectrometers are used at synchrotrons and XFEL facilities [9,10].

In addition, it has been recently shown that HEROS spectra can be measured shot-by-shot at XFEL sources to observe the variation of an absorption spectrum on very short time scales. This will possibly allow, thanks to MOSARIX spectrometer, following a chemical process in real time [3].

## Photons/electrons coincidence measurements

The coincidence measurement of energy-resolved-photons with ions or electrons will open new horizons to access fundamental atomic and molecular information.

Following inner-shell ionization, the highly excited ion can decay by x-ray fluorescence or by ejection of an Auger electron. Recent X-ray photon-ion coincidences without resolving the photon energy on Argon have revealed the complexity of the interplay between these competing electronic decay processes[11]. X-ray fluorescence is favoured for decay of deep excited inner shells in heavy atoms, while Auger decay is favoured for light elements and shallow excited inner-shells. After X-ray emission ($K_{\alpha, \beta...}$) subsequent Auger decays (e.g. LMM, LMN, MNN, etc,…) occur. By detecting in coincidence X-ray photons with all subsequent electrons, thanks to a magnetic bottle time-of-flight spectrometer (MB-TOF) with a very high detection efficiency (up to 90%), it will be feasible to disentangle all decay channels involving multiple Auger decay (direct and cascade). Owing to the good detection efficiency of the MOSARIX spectrometer, this coincidence measurement will become possible. The ~1ns time response of the photon detector is necessary to allow electron time of flight determination in the MB-TOF and can be used in multi-bunch operation of the synchrotron while the detection of electrons is restricted to single bunch operation to define the time zero.

A much more challenging experiment would be to use photon-electron coincidence to gain an insight in radiative Auger decay when the energy is shared between an X-ray photon and an electron. Since this is a minor process, it will be difficult to extract, but the good energy resolution of the MOSARIX spectrometer and the high efficiency of the MB-TOF are the only hope to observe this process by considering the energy balance between the photon and the electron.

In the case of molecules, ultrafast dissociation may take place in the course of cascade of relaxation following the deep electron vacancies creation. Multistep ultrafast dissociation is induced by the promotion of a deep inner-shell electron to an antibonding molecular orbital. Following the excitation, the system relaxes via a series of subsequent radiative or non-radiative (i.e., Auger) decay steps. Recently, Travnikova et al.[12,13] found out that tuning hard x-ray excitation energy along Cl 1s → σ* resonance in gaseous HCl allows manipulating molecular fragmentation in the course of the induced multistep ultrafast dissociation.

## FISIC (ASUR)

X-ray spectroscopy has also proved to be a very powerful tool to investigate quantum dynamics in heavy ions collisions with matter of whatever nature, dilute or condensed[5,14]. With the advent of the new-generation facilities such as S3/SPIRAL2/GANIL[1] in Caen and CRYRING/FAIR[2] in Darmstadt (Germany), research opportunities in the domain of fast (MeV/u) ion- slow (keV/u) ion collisions are open, offering the possibility to study the fundamental electronic processes in a hitherto unexplored collision regime. The measurement of photons (resolved in energy) in coincidence with the ions brings then information on the excitation processes.

---

[1] http://pro.ganil-spiral2.eu/spiral2/instrumentation/s3/equipex-s3
[2] https://www.gsi.de/work/forschung/appamml/atomphysik/anlagen_und_experimente/cryringesr.htm

Knowledge of fundamental electronic mechanisms at play in ion-ion collisions can provide a real breakthrough in the understanding of energy transfer in various plasmas such as inertial confinement fusion plasma, stellar/interstellar plasmas and also in material damages. The Fast Ion (MeV/u) –Slow Ion (keV) Collision project (FISIC) aims at measuring absolute electronic cross sections in the intermediate velocity regime, a regime in which the ion sopping power is maximum and where all the primary electronic processes (electron capture, loss and excitation) reach their optimum. In this regime, there is a clear lack of measurements, and available theoretical calculations are at their limit of validity. In other words, this regime corresponds to a real *"terra incognita"* for atomic physics. To investigate deeply the N-body quantum dynamics in ion-matter interaction, we will need to reach the pure three-body problem (bare ion on hydrogenic target) as a benchmark and then explore the role of additional electrons bounded to the target and/or to the projectile -one by one- to quantify several effects such as: electron-electron interactions, multi-electron processes and Coulomb forces acting on the electron cloud in the entrance and exit pathways of the collision. The high-energy ion beams will be provided by the new facilities such as S3/SPIRAL2/GANIL and CRYRING/FAIR/GSI while the low-energy ion beams are produced by an ECR ion source equipped with a specific beam line to get a full control on the ion beam shape and on the charge state. With FISIC, we intend to build a unique crossed beam ion–ion experiment to perform absolute cross section measurements. Among the most severe bottlenecks that should be overcome, we can stress the resistance of the target for stripping of the high-energy ion beam; the control of the overlap of the crossed beams and the mastership of the overall experimental conditions to optimize the multi-coincidence event detection. To measure the capture and ionisation processes (with cross sections excepted to be in the $10^{-16} - 10^{-18}$ cm² with ions of atomic number $\leq 18$), charge-state analysers equipped with ion detectors having a time resolution of ~1 ns are under development. For instance, for the high-energy ions, new prototypes as YAP:Ce scintillators and diamond-based detectors are currently tested in collaboration with our German colleagues at GSI and Jena. To obtain information on the formation of excited states during the ion-ion collision, the measurement of the radiative decay of excited ions is required, implying the use of a high-transmission high-resolution x-ray spectrometer. To give an example, for $Ar^{18+}$ (@ 8MeV/u) + $Ar^{16+}$ (@ a few keV/u)→ $Ar^{18+}$ + $Ar^{16+*}$ → $Ar^{18+}$ + $Ar^{16+}$+ (~3keV) $h\nu$, the excitation cross section is expected to be at maximum ~$10^{-18}$ cm² according to the most available sophisticated calculations. In many cases, the value to be measured could be much lower. With the FISIC program, a wide range of collision systems will be investigated, from symmetric collisions (Zp ~ Zt; where Zp is the atomic number of the fast ion and Zt the one of the slow ion) reachable at S3/SPIRAL2 to asymmetric collisions (Zp > Zt and even Zp >> Zt) accessible at CRYRING/FAIR.

## II. State f-the-art x-ray emission spectrometers

Owing to the availability of high-resolution x-ray position-sensitive detectors, there has been a fast recent development of Von Hamos x-ray emission spectrometers worldwide, providing excellent compromise between resolution and efficiency [15]. Several methods have been used in order to increase the efficiency by using larger crystal solid angle. Alonso-Mori and coworkers installed up to 16 crystals[9]. Szlachetko and coworkers chose the option of a large segmented crystal [10]. A more exotic geometry using full-cylindrical crystal has been recently built[16]. All of these von Hamos spectrometers are optimized for x-ray emission spectroscopy between 5 and 10 keV. The other option to maximize the efficiency of the spectrometer is to use mosaic crystals having higher reflectivity. This is the case of the recent implementation of graphite mosaic crystals in von Hamos geometry spectrometer to measure X-ray in the range 4.5-10 keV [17,18] and 8 - 60 keV [19] with a resolution power < 2000.

The groups involved in this project have a strong interest in the 2-5 keV photon energy range because it covers the K emission lines of sulphur, chlorine and argon as well as the L emission lines of iodine and xenon. The 2-5 keV photon energy range is challenging because conventional silicon crystals cannot be used and because there is a strong absorption in air in this energy domain. We found an interesting von Hamos spectrometer in this energy domain[20].

For the design of MOSARIX, we used Alonso-Mori's idea of a multi-crystal geometry with mosaic HAPG crystals in order of to gain one order of magnitude in the efficiency.

## III. Organization of the project

**Project leader:** Marc Simon.

**Steering Committee:** Marc Simon, Dominique Vernhet and Francis Penent.

Project manager: Iyas Ismail (P@P CDD until 12/2018).

The whole **consortium** is composed by the following members:

**Laboratoire de Chimie Physique-Matière et Rayonnement (LCPMR)** : Marc Simon (DR), Loïc Journel (Pr), Tatiana Marchenko (CR), Renaud Guillemin (CR), Thierry Marin (IE), Oksana Travnikova (CR), Francis Penent (DR), Pascal Lablanquie (DR), Jérôme Palaudoux (MCF), Lidija Andric (MCF) and Régis Vacheresse (IE)

**Institut des Nanosciences de Paris (INSP)** : Emily Lamour (Pr), Anna Lévy (MCF), Stéphane Macé (IE), Christophe Prigent (MCF), Jean-Pierre Rozet (Pr. Em.), Sébastien Steydli (IE), Martino Trassinelli (CR), and Dominique Vernhet (DR).

The whole expenses process, the assembly, the high demanding control-command software, the first test and the superposition of up to 8 images have to be realized. This will require a precise alignment of the different crystals.

The project manager (I. Ismail) has obtained a CNRS permanent position at LCPMR and leaves his responsibilities in the MOSARIX project starting 01 December 2018. The continuity of the work requires the temporary recruitment of an engineer (IE) for 2 years. The recruited engineer will implement the design of the spectrometer and will write the control and acquisition software of the spectrometer. He/she will also participate in the design and the construction of a 4-crystals version more adapted to the needs of INSP team. This work will be performed with the guidance of Iyas Ismail and Loïc Journel.

**Supprimé:** Ingénieur de Recherches (IR)

## IV. Development of MOSARIX

### Performed works

**Simulations and crystal choice**

Ray-tracing simulation has been performed to evaluate the possibility of using curved mosaic crystals in von Hamos geometry. The results of this simulation showed that energy resolution is mainly dominated by the broadenings due to both the detector spatial resolution and the size of the source. The broadening caused by the mosaic-spread of a curved crystal was found to be very similar to that of a plain crystal. It should be noted that the broadening associated to the stress induced in the lattice planes ($\Delta E_{bend}$) caused by the crystal bending has not been considered in this simulation.

The mosaic graphite crystal has been selected as our best choice of crystals because of its high integrated reflectivity. Furthermore, it allows working at the first order of reflection in the energy range 2-5 keV. Among mosaic graphite crystal, the novel Highly Annealed Pyrolytic Graphite (HAPG) has been chosen since it can be produced with lower mosaic spread (about 0.05°) compared to the other most-commonly used mosaic Highly Oriented Pyrolytic Graphite (HOPG) crystal (typically 0.4 - 1.2°). HAPG crystal is fabricated as thin flexible sheets of desired thickness that can be deposited on substrates of almost any shape (see for instance [L. Anklamm et al.[21]). This is possible, according to the manufacturer, because $\Delta E_{bend}$ is negligible comparing to the others broadening sources. However, we decided to verify this assumption experimentally by measuring the overall resolution obtained with this crystal.

This choice of HAPG crystal instead of HOPG allows higher energy resolution and minimizes the focal aberrations. HAPG has been used in von Hamos geometry to measure photons in the range 4.5-10 keV [17,18] and 8 - 60 keV [19] but has not been tested yet below 4.5 keV. This is why we have concluded a deal with Optigraph GmbH Company (exclusive manufacturer of HAPG) to provide us a customized HAPG crystal to test the resolution and reflectivity in our energy range.

**Fast-time and position-sensitive x-ray detector**

In the absence of commercially-available detectors that fulfil our requirements of fast-time and position-sensitive detection of photons in coincidence mode, we have developed a new micro-channel-plates (MCPs) based detector for keV energy photons. The concept of this detector consists of the pairing of the MCPs with a multi-layers photocathode. This allows enhancing the efficiency (up to 30%) without degrading neither the spatial (100 μm) nor the time resolution (100 ps). This detector was the subject of a French patent application (N° FR1850256, application January 12 2018) and a publication by Ismail *et al.*[22].

**One-crystal spectrometer**

In order to validate the simulations and to measure the resolution and the reflectivity of the HAPG crystal in the 2-5 keV energy range, we have built a prototype version of the spectrometer that consists of one curved HAPG crystal and our new MCPs-based detector. A curved HAPG crystal of a = 25 mm along the dispersive direction and b = 110 mm (in the non-dispersive direction) with a radius of curvature R=500 mm has been used. The substrate is made of N-BK7 glass and coated with HAPG layer (40 μm thickness) by the company Optigraph. This spectrometer, working under a vacuum of $10^{-6}$ mbar, was commissioned initially at our laboratory and then during two beamtimes (May and September 2017) at GALAXIES beamline of SOLEIL synchrotron. The results obtained with this spectrometer were very encouraging. A resolving power of E/ΔE=4000 @ $\Theta_B$ =38.67° was obtained by measuring the elastic peak and was also verified by measuring potassium Kα emission spectrum. The measured integrated reflectivity of the HAPG crystal (*Ri* = 1.4 ± 0.28 mrad) is in very good agreement with theoretical XOP calculations[23].

**Multi-crystals spectrometer**

The successful operation of the one-crystal spectrometer encouraged us to go toward the multi-crystals version of the spectrometer. Several designs of the spectrometer have been studied.

All the crystals would have the same characteristics to the single-crystal prototype: 500 mm curvature, 25 * 110 mm$^2$.

One of the challenges will be to obtain several superimposable images for each crystal.

**In-vacuum version**

Given that most of the experiments in which MOSARIX will be used operate in-vacuum, the intuitive choice was to opt for an in-vacuum operating spectrometer. This choice is also convenient for the MCPs-based detector. Two in-vacuum operating configurations have been studied.

*"Translation" configuration*

This solution is classically used for spectrometers working in von Hamos geometry [for example: Alonso-Mori et al,[9] and J. Szlachetko et al. [10]. It consists of mounting the detector and the crystals holder on two parallel translation stages. The movement of these translations allows changing the Bragg angle. The distance between the transitions is equal to the radius of curvature of the crystal, in order to satisfy the focusing condition (source and detector are in the focal axis of the crystal).

This configuration requires the use of a huge vacuum chamber of 3 x 1x 1 m$^3$ to house both the crystals and the detector, and to cover the energy range of the spectrometer. The weight of such chamber is estimated to be more than 1.3 tonnes. This solution might be a good choice for a permanent end-station installation but is not convenient for a transportable spectrometer mainly because of the non-manoeuvrable and heavy chamber.

*"Rotation" configuration*

This second solution allows selecting the Bragg angle using two rotation stages (Θ, 2Θ) on which the crystals holder and the detector are mounted. The crystals are placed inside a vacuum chamber equipped with two vacuum tubes connecting it to both the source (chamber) and the detector. For each Bragg angle, the positions of the detector and crystals-holder must be adjusted to satisfy the focusing condition governed by the radius of curvature of the crystal. This adjustment requires the using of sets of vacuum tubes of different lengths for each (central) Bragg angle.

This solution has the advantage of being smaller and less heavy (only the crystals are inside the vacuum chamber) than the "translation" configuration. However, it requires using a set of tubes per "main" energy, which presents a major limitation in case one needs to look at different energies (e.g. different elements) during an experiment. This will require changing the tubes, which would cause an important loss of time to vent, re-align and then re-pump the setup.

**In-air version**

The limitations of the in-vacuum configurations in addition to the very recent development of a new x-ray hybrid detector (TimePix3) incited us to look for an alternative design.

TimePix3 is a time- and position-sensitive x-ray detector (resolution: 50 µm and 1.25 ns, high count-rate of 40 M hits/s) developed by the CERN and commercialized in October 2017 by ADVACAM Company.

TimePix3 is designed to detect higher-energy photons and has not yet been used so far below 5 keV. We arranged a deal with ADVACAM Company to test this detector in our laboratory at lower energies. Our test performed at LCPMR revealed that TimePix3 allows the detection of photons in the 2-5 keV energy with efficiency between 20 and 30 %. This detector fulfils our requirement of coincidence measurement with the great advantage of both in-air and in-vacuum (tested at 10$^{-6}$ mbar) working capability. With this detector, we could consider a new design of the spectrometer working in-air (with He balloons).

## Ongoing & scheduled works

This in-air design (fig.1) is much lighter, transportable and flexible compared to the previous designs. The spectrometer will be fully motorized and controlled by PC. The final version is designed to include 48 picomotors for the alignment of the 16 crystals (3 per crystal). The ensemble of the crystals will be mounted on a rotation stage (Θ) while the detector (TimePIX3) will be mounted on a second rotation stage (2Θ) having the same rotation axis as the first stage. Two translation stages will insure the adjustment of the distances source-crystal and crystal-detector automatically and rapidly for each Bragg angle. The detector will be mounted on an additional rotation stage to place it parallel to the dispersion axis of the crystal.

This design has been realized with the active implication of Mr. Régis VACHERESSE (LCPMR). Advanced negotiations are engaged with Newport Company for the construction of the "delicate parts" (mainly the heavy rotation stages) of the spectrometer and with PI Company for the motorization of the crystals. Many parts will be produced by the mechanical workshop of the LCPMR.

Although the design is made for 16 crystals, we are considering using only a reduced number of crystals (probably 8) as a first step. The remaining crystals (and their motorizations) would be ordered later on, however their implementation on the spectrometer is already considered so it will be done without much difficulty.

We also wish to build a second version of the spectrometer operating under-helium with 4 crystals more adapted to the needs to the INSP team to work at the vicinity of 2.9 keV.

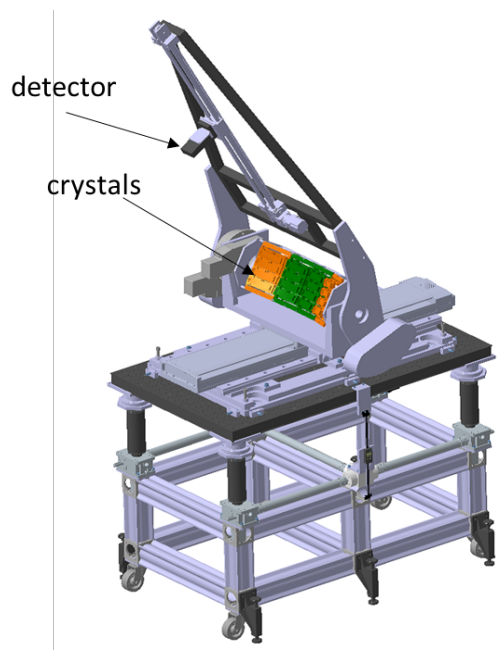

FIG.1 The *in-air* design of MOSARIX. The implementation of the chassis is not yet completed.

A commercially available large x-ray pressure window (from Luxel Company) can be mounted on the exit flange of the interaction vacuum chamber for the outpu of the photons from the vacuum toward the spectrometer (in air). According to the manufacturer, this window has a He leak rate <$10^{-9}$ mbar L/s which allows using it with UHV chamber. Moreover, a more conservative solution can be considered with a differential pumping scheme by using two of these windows together mounted on a UHV gate valve equipped with a pumping exit.

The crystals and the detector will be mounted in an He-filled balloon in order to reduce the attenuation of x-ray photons.

The negotiations to command the crystals and customized version of the TimePix3 are completed. We also obtained a discount of 25% (33 k€) for the purchase of 16 crystals from Optigraph GmbH Company.

## V. Choice of the detector

The final choice of the detector has not yet been finalized.

TimePix3 has the advantage to allow coincidence/covariance mapping measurements with a relatively low detection efficiency (between 20 and 30%).

Conventional CCD camera has the advantage to have a much higher detection efficiency, up to more than 90% but coincidences are then impossible to perform.

Since coincidence measurements will be very challenging with the detection efficiency provided by TimePic3, we plan to start the project with a CCD camera.

## VI. Specifications of MOSARIX

- Photon energy range: 2 to 5 keV @$1^{st}$ order of reflection.
- Resolution < 1 eV (achieved with one crystal – to be realized with 8 crystals)
- Fast detector to allow coincidence measurements.
- Global efficiency $10^{-5}$ with 16 crystals depending on the energy and on the detector.
- 72 eV recorded in parallel at 3 keV of photon energy

## VII. Financial point

No additional budget is asked (initial budget 400 k€) to realize MOSARIX.

- Global budget **400 k€.**
- Spent (until July 2018) **85 k€.**

- Available **315 k€.**

The scheduled spending are:

- Mechanical & motorization parts of the spectrometer **100 k€.**
- Detector: TimePix3 **60 k€** (for information: MCPs-based detector 90 k€).
- Vacuum widow(s) setup + differential pumping **35 k€**.
- 16 crystals **100 k€:**

  One HAPG crystal 8250€, x16= 132 k€.
  33 k€ special discount of 25% for the purchase of 16 crystals.

- *If it is decided to use only 8 crystals, the saved budget (≈ 40k€) + the remaining 20 k€ (=60 k€) will be used to build a second version of the spectrometer (under vacuum with 4 crystals) or to buy a CCD camera.*

## VIII. Attractiveness of MOSARIX

Several groups worldwide have expressed their interest in the development of MOSARIX. At LCPMR, Philippe JONNARD (DR) is interested to perform Resonant Inelastic X-ray Scattering on multilayers. Jean-Pascal RUEFF, beamline manager of the SOLEIL-GALAXIES beamline is willing to use this spectrometer because his diffractometer is not able to perform any measurements below 5 keV of photon energy. Matjaz ZITNIK and his group (Ljubljana) are particularly interested by the photon-electron coincidences. Philippe Wernet (Uppsala) is wishing to perform RIXS on 4d transition metal complexes.

## IX. Schedule for the first measurements

The first measurements will be performed at SOLEIL on the GALAXIES beamline in 2020 and at the European XFEL at the SQS beamline in 2021. The FISIC measurements will start in 2023 when the SPIRAL II beam will be available.

**References**

1. Rudek, B., Son, S., Rolles, D. & Al., E. Ultra-efficient ionization of heavy atoms by intense X-ray free-electron laser pulse. *Nat. Photonics* **6,** 858–865 (2012).

2. Marchenko, T. *et al.* Electron Dynamics in the Core-Excited $CS_2$ Molecule Revealed through Resonant Inelastic X-Ray Scattering Spectroscopy. *Phys. Rev. X* **5,** 031021 (2015).

3. Kavcic M *et al.* Hard x-ray absorption spectroscopy for pulsed sources. *Phys. Rev. B* **87,** 075106 (2013).

4. Błachucki, W. *et al.* High Energy Resolution Off-Resonant Spectroscopy for X-Ray Absorption Spectra Free of Self-Absorption Effects. *Phys. Rev. Lett.* **112,** 173003 (2014).